\documentclass[runningheads]{llncs}

\usepackage{eccv}

\usepackage{eccvabbrv}

\usepackage{graphicx}
\usepackage{booktabs}

\usepackage[accsupp]{axessibility}  

\usepackage{textcomp}
\usepackage{newunicodechar}
\newunicodechar{−}{\textminus}

\usepackage{hyperref}

\usepackage{orcidlink}

\usepackage{multirow}
\begin{document}

\title{SALIENT: Frequency-Aware Paired Diffusion for Controllable Long-Tail CT Detection\thanks{
Code and trained models are planned for public release under terms permitting non-commercial academic research use. Final licensing details will be specified at the time of release.
}}





\author{
Yifan Li\inst{1,2}
\and Mehrdad Salimitari\inst{1,2}
\and Taiyu Zhang\inst{1,2}
\and Guang Li\inst{2}
\and David Dreizin\inst{1,2}
}

\authorrunning{Y. Li et al.}

\institute{
\inst{1} Trauma Radiology AI Laboratory (TRAIL),\\
Department of Radiology and Nuclear Medicine,\\
University of Maryland School of Medicine
\and
\inst{2} Department of Radiology and Nuclear Medicine,\\
University of Maryland School of Medicine
}

\maketitle

\begin{abstract}
Detection of rare lesions in whole-body CT is fundamentally limited by extreme class imbalance and low target-to-volume ratios, producing precision collapse despite high AUROC. Synthetic augmentation with diffusion models offers promise, yet pixel-space diffusion is computationally expensive, and existing mask-conditioned approaches lack controllable attribute-level regulation and paired supervision for accountable training. We introduce SALIENT, a mask-conditioned wavelet-domain diffusion framework that synthesizes paired lesion–mask volumes for controllable CT augmentation under long-tail regimes. Instead of denoising in pixel space, SALIENT performs structured diffusion over discrete wavelet coefficients, explicitly separating low-frequency brightness from high-frequency structural detail. Learnable frequency-aware objectives disentangle target and background attributes (structure, contrast, edge fidelity), enabling interpretable and stable optimization. A 3D VAE generates diverse volumetric lesion masks, and a semi-supervised teacher produces paired slice-level pseudo-labels for downstream mask-guided detection. SALIENT improves generative realism, as reflected by higher MS-SSIM (0.63$\rightarrow$0.83) and lower FID (118.4$\rightarrow$46.5).In a separate downstream evaluation, SALIENT-augmented training improves long-tail detection performance, yielding disproportionate AUPRC gains across low prevalences and target-to-volume ratios. Optimal synthetic ratios shift from $2\times$ to $4\times$ as labeled seed size decreases, indicating a seed-dependent augmentation regime under low-label conditions. SALIENT demonstrates that frequency-aware diffusion enables controllable, computationally efficient precision rescue in long-tail CT detection.
\end{abstract}

\section{Introduction}

Whole-body CT (WBCT) is widely used for cancer staging, inflammatory disease evaluation, and polytrauma assessment \cite{vorontsov2019deep, huang2024lidia, roth2022rapid, dreizin2020multiscale, dreizin2020deep}. Despite advances in deep learning, detection of uncommon or small lesions in WBCT remains fundamentally difficult. Two compounding failure modes drive this challenge. First, within-patient signal dilution arises from low target-to-volume ratios (TVRs) in large torso fields of view \cite{lee2021clinical, dreizin2023artificial}. Second, cross-dataset prevalence dilution produces extreme class imbalance in long-tail detection settings \cite{dreizin2012blunt, dreizin2015multidetector}. Together, these effects create a precision ceiling that cannot be resolved solely through architectural modification \cite{oakden2020hidden, daneshjou2022disparities, salmi2024handling}.

Segmentation networks such as nnU-Net \cite{isensee2021nnu} and hybrid CNN–Transformer architectures \cite{roy2023mednext, woo2023convnext} have achieved strong voxel-level performance. However, lesion detection under severe imbalance remains sensitive to background dominance. Even when AUROC appears high, models frequently suffer from poor precision, low AUPRC, and unstable F1 scores \cite{salmi2024handling, hasani2022artificial}. In deployment, low precision leads to spurious saliency and increased false alarms, limiting clinical trust and usability \cite{bernstein2023can, katal2024ai}. Attention mechanisms and mask-guided training can improve feature localization \cite{li2018tell, wu2024development, arrieta2020explainable, yan2024enhanced}, yet they do not address the underlying scarcity of informative positive samples.

Synthetic data augmentation has long been proposed to mitigate data sparsity \cite{langlotz2019roadmap}. Diffusion probabilistic models (DDPMs) have recently surpassed GANs in image quality metrics for high-dimensional medical imaging \cite{khader2023denoising, muller2023multimodal}. However, existing mask-conditioned diffusion approaches often rely on fixed input geometries or limited sources of structural diversity \cite{dorjsembe2024conditional, heo2025controllable, zhang2023adding}. Pixel-space DDPMs are computationally prohibitive in 3D and often require aggressive downsampling, degrading fine-grained in-plane detail critical for small lesion modeling. Frequency-domain diffusion methods offer computational speedups \cite{phung2023wavelet}, but rely on manually tuned band weights and do not disentangle interpretable image attributes such as brightness, structure, and detail.

Moreover, augmentation is typically assumed to provide monotonic benefit \cite{khosravi2024synthetically, prakash2025evaluating, cha2020evaluation}, yet augmentation dose-response behavior remains uncharacterized. We are not aware of prior work that defines optimal ("therapeutic dose") synthetic augmentation levels or identifies performance degradation under excessive synthetic sampling (“toxic doses”) in long-tail detection \cite{johnson2019survey}. Consequently, synthetic augmentation remains heuristic rather than prescriptive.

In this work, we introduce SALIENT (Structured Attention-Leveraged Inference for Edge-aware Neural Training), a mask-conditioned wavelet-domain diffusion framework for controllable CT augmentation. SALIENT replaces pixel-space denoising with structured diffusion over discrete wavelet coefficients, explicitly separating global brightness from high-frequency detail. We introduce learnable frequency-domain weighting that disentangles target and background attributes into interpretable optimization “dials” governing structure, detail, contrast, brightness, and spatial context. This design yields substantial computational speedup while preserving high-resolution boundaries essential for small lesions.

\begin{figure*}[t]
    \centering
    \includegraphics[width=\textwidth]{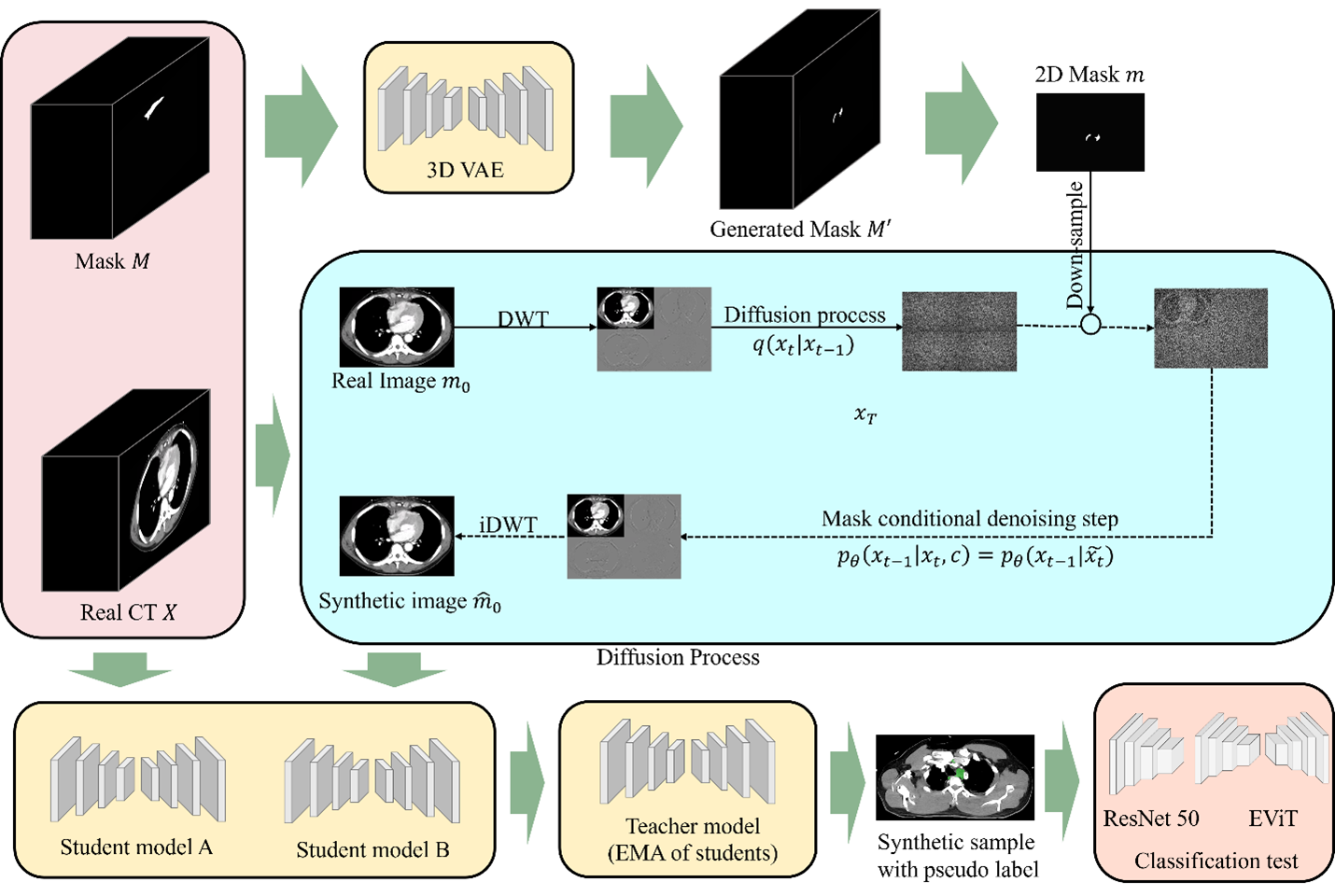}
    \caption{
    Overview of the proposed SALIENT synthetic data and classification pipeline. 
    Real CT volumes and lesion masks are first processed by a 3D VAE to generate diverse volumetric masks, which are projected into 2D slice space and used as conditioning signals for the wavelet-domain diffusion model. 
    SALIENT operates on discrete wavelet coefficients to synthesize mask-guided CT slices, which are subsequently pseudo-labeled by a semi-supervised segmentation teacher (UCMT). 
    The resulting synthetic CT--mask pairs augment training for a slice-level mask-guided ResNet-50 classifier. Slice-level predictions are further aggregated into subject-level decisions using an Embedded Vision Transformer (EViT)\cite{islam2024seeking}.
    }
    \label{fig:pipeline_overview}
\end{figure*}

Crucially, SALIENT generates anatomically coherent lesion–mask pairs sampled from a learned latent pathology manifold, enabling direct training of mask-guided detectors. We further characterize the \emph{augmentation dose-response} under varying prevalence and seed sizes, revealing a stable therapeutic regime and a rightward dose shift under low-label conditions. These findings suggest that frequency-aware diffusion enables controllable precision rescue in long-tail detection settings.

Our contributions are:

\begin{itemize}
\item A mask-conditioned wavelet-domain diffusion framework with learnable frequency weighting for attribute-specific control.
\item Paired synthetic lesion–mask generation enabling accountable mask-guided detection training.
\item Empirical characterization of augmentation dose-response behavior under varying prevalence and labeled seed sizes.
\item A practical foundation for archetype-guided augmentation scaling in long-tail CT detection.
\end{itemize}

\section{Related Work}

\subsection{Long-Tail Detection in Medical Imaging}

Detection of rare or small targets in cross-sectional imaging is challenged by extreme class imbalance and low target-to-volume ratios \cite{lee2021clinical, dreizin2023artificial, salmi2024handling}. While voxel-level segmentation networks such as nnU-Net \cite{isensee2021nnu} and modern CNN–Transformer hybrids \cite{roy2023mednext, woo2023convnext} achieve strong segmentation performance, detection precision under imbalance remains limited. Attention-based and mask-guided paradigms improve feature localization and accountability \cite{li2018tell, wu2024development, arrieta2020explainable, yan2024enhanced}, but do not fundamentally address the scarcity of informative positive samples.

\subsection{Synthetic Augmentation with Diffusion Models}

Generative augmentation has been widely explored to address data sparsity \cite{langlotz2019roadmap}. GAN-based approaches were historically dominant but suffered from training instability and mode collapse. Diffusion probabilistic models (DDPMs) have demonstrated superior perceptual realism and stability in medical imaging \cite{khader2023denoising, muller2023multimodal}. Mask-conditioned diffusion variants allow for structural control \cite{dorjsembe2024conditional, heo2025controllable, zhang2023adding}, yet most approaches bind lesion geometry to conditioning masks or preserve the surrounding background, which limits morphological diversity.

Pixel-space diffusion in 3D is computationally expensive and often requires reduced spatial resolution. Latent-space diffusion reduces cost but may obscure fine-grained structural detail. Recent frequency-domain diffusion methods introduce spectral acceleration \cite{phung2023wavelet}, yet typically rely on manually tuned band weights without interpretable control over image attributes.

In contrast, SALIENT integrates mask conditioning with wavelet-domain diffusion and learnable attribute-specific frequency weighting. This enables explicit disentanglement of target and background structure, detail, contrast, and brightness, providing interpretable and task-aligned generative control.

\subsection{Augmentation Dose-Response}

Most augmentation studies assume monotonic performance gains with increasing synthetic data \cite{khosravi2024synthetically, prakash2025evaluating, cha2020evaluation}. However, oversampling is known to risk overfitting and degraded generalization in imbalanced learning \cite{johnson2019survey}. Systematic characterization of augmentation dose-response curves, therapeutic regimes, or toxicity effects remains underexplored. No existing framework provides prescriptive guidance for optimal augmentation under varying prevalence or labeled seed sizes.

We address this gap by empirically quantifying augmentation dose-response behavior under controlled prevalence and seed conditions, demonstrating a reproducible therapeutic regime and a seed-dependent dose shift.

\section{Proposed Method}

We propose an end-to-end pipeline for controllable CT augmentation and long-tail detection (Fig.~\ref{fig:pipeline_overview}). SALIENT generates paired synthetic CT--mask samples via wavelet-domain diffusion conditioned on VAE\cite{pinheiro2021variational}-sampled lesion masks, and performance is evaluated using mask-guided slice-level classification with subject-level aggregation.

\subsection{Dataset}

The dataset comprises contrast-enhanced whole-body CT examinations from an adult trauma cohort. Mediastinal hematoma occurs in approximately 3\% of studies, yielding a naturally long-tail detection problem characterized by low target-to-volume ratios and irregular, multiscale morphology. Mediastinal hematomas are potentially clinically significant, making this a practical stress-test for precision under signal dilution. We evaluate on an internal contrast-enhanced torso CT cohort comprising 5,205 subjects (200 mediastinal hematoma–positive, 5,005 negative controls) following IRB approval. Positive subjects were confirmed by board-certified radiologists using clinical reports and dedicated image review. All scans underwent unified preprocessing, including reorientation to a consistent coordinate convention, resampling to common voxel spacing, soft-tissue HU windowing, intensity normalization, and anatomical cropping to the mediastinal region of interest. All experiments are conducted on the mediastinal hematoma task under controlled prevalence shifts to simulate long-tail detection regimes.

\subsection{SALIENT}

\begin{figure*}[t]
    \centering
    \includegraphics[width=\textwidth]{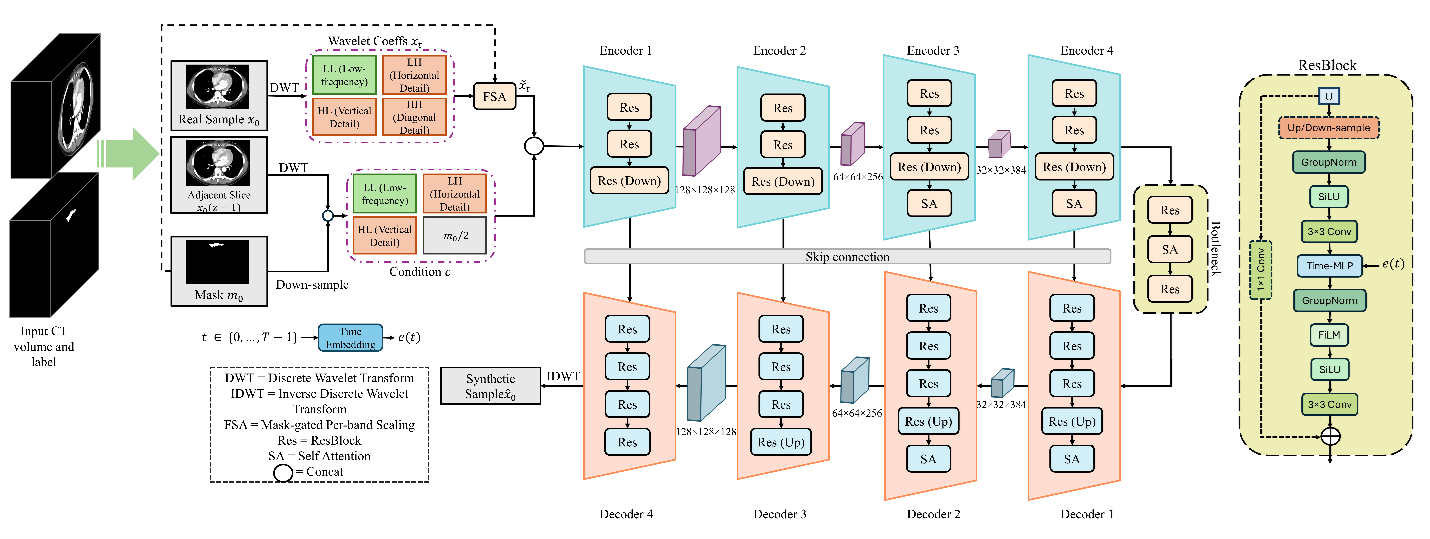}
    \caption{
    Architecture of SALIENT in the wavelet domain. 
    A central CT slice and its axial neighbors are transformed into wavelet coefficients (LL, LH, HL, HH). 
    A mask-gated frequency scaling (FSA) module modulates the noisy coefficients before concatenation with a 2.5D conditioning stack. 
    A time-conditioned UNet predicts clean wavelet coefficients at each diffusion step, which are reconstructed into synthetic CT slices via inverse DWT.
    }
    \label{fig:architecture}
\end{figure*}

\subsubsection{Overview.}
As illustrated in Fig.~\ref{fig:architecture}, SALIENT is a wavelet-domain conditional diffusion model designed to generate anatomically coherent and mask-consistent CT slices. 
Rather than operating directly in pixel space, the model performs diffusion over discrete wavelet coefficients, enabling explicit control over low-frequency structure and high-frequency detail. 
Mask and 2.5D anatomical context are incorporated both architecturally and through structured guidance during sampling.

\subsubsection{Wavelet-Domain Formulation.}
Let $X \in \mathbb{R}^{Z \times H \times W}$ denote a preprocessed CT volume and $m_z \in \{0,1\}^{H \times W}$ the lesion mask at slice $z$. 
For each central slice $x_z \in \mathbb{R}^{H \times W}$, we define a 2.5D neighborhood
\[
\mathcal{N}(z) = \{ x_{z+\Delta} : \Delta \in \mathcal{S} \},
\]
where $\mathcal{S}$ contains small axial offsets capturing through-plane continuity.

We apply a single-level Haar discrete wavelet transform:
\begin{equation}
w_z = \mathcal{W}(x_z)
= [LL_z, LH_z, HL_z, HH_z]
\in \mathbb{R}^{4 \times \frac{H}{2} \times \frac{W}{2}},
\end{equation}
where $LL$ encodes low-frequency structure and $LH$, $HL$, $HH$ capture oriented high-frequency components.

We learn a conditional diffusion model
\begin{equation}
p_\theta(w_z \mid c_z),
\end{equation}
where $c_z$ encodes the down-sampled central mask and selected neighbor wavelet bands.

The forward process follows
\begin{equation}
q(w_t \mid w_0)
=
\mathcal{N}
\left(
w_t;
\sqrt{\bar{\alpha}_t} w_0,
(1-\bar{\alpha}_t)\mathbf{I}
\right),
\end{equation}
and the reverse model predicts clean coefficients:
\begin{equation}
\hat{w}_0 = f_\theta(w_t, t, c_z).
\end{equation}

Synthetic CT slices are reconstructed via
\begin{equation}
x_z^{\text{syn}} = \mathcal{W}^{-1}(\hat{w}_0).
\end{equation}

\subsubsection{Mask-Guided Wavelet UNet.}
At timestep $t$, noisy coefficients
\[
w_t \in \mathbb{R}^{4 \times \frac{H}{2} \times \frac{W}{2}}
\]
are first modulated by a mask-gated frequency scaling (FSA) module to produce $\tilde{w}_t$. 
The conditioning tensor
\[
c_z \in \mathbb{R}^{C_{\text{cond}} \times \frac{H}{2} \times \frac{W}{2}}
\]
includes the down-sampled mask and selected neighbor wavelet bands. 
The UNet input is
\begin{equation}
u_t = [\tilde{w}_t \| c_z].
\end{equation}

The backbone is a four-level encoder--decoder UNet with residual blocks, symmetric skip connections, and self-attention at the deepest resolutions. Diffusion timesteps are encoded via sinusoidal embeddings and injected through FiLM-style scale–shift conditioning.

\subsubsection{Objective and Modeling Principle}

SALIENT is a mask-conditioned wavelet-domain diffusion model designed to generate anatomically coherent CT slices and paired lesion masks under extreme class imbalance. 

Let $x_z \in \mathbb{R}^{H \times W}$ denote a central CT slice and $m_z \in \{0,1\}^{H \times W}$ its lesion mask. Instead of modeling pixel intensities directly, we operate in the discrete wavelet domain:
\begin{equation}
w_z = \mathcal{W}(x_z) = [LL_z, LH_z, HL_z, HH_z] \in \mathbb{R}^{4 \times \frac{H}{2} \times \frac{W}{2}},
\end{equation}
where $LL$ captures global structure and brightness, and $LH/HL/HH$ encode oriented high-frequency detail.

Our goal is to learn a conditional generative model:
\begin{equation}
p_\theta(w_z \mid c_z),
\end{equation}
where $c_z$ includes the down-sampled lesion mask and selected neighboring wavelet features (2.5D context). 

The forward diffusion process follows:
\begin{equation}
q(w_t \mid w_0) =
\mathcal{N}(\sqrt{\bar{\alpha}_t} w_0, (1-\bar{\alpha}_t)I),
\end{equation}
and the reverse model predicts clean coefficients:
\begin{equation}
\hat{w}_0 = f_\theta(w_t, t, c_z).
\end{equation}

Synthetic slices are reconstructed via inverse wavelet transform:
\begin{equation}
x_z^{syn} = \mathcal{W}^{-1}(\hat{w}_0).
\end{equation}

Operating in wavelet space provides two advantages:  
(i) explicit separation of global brightness (LL) from high-frequency boundaries,  
(ii) structured control over target–background detail without full 3D pixel-space diffusion.

\subsubsection{Wavelet-Aware Training Objective}

Unlike standard diffusion trained with uniform $\ell_2$ loss, SALIENT optimizes a frequency-structured objective that disentangles global structure from boundary detail.

We define a band-weighted reconstruction:
\begin{equation}
\mathcal{L}_{wavelet}
=
\mathbb{E}
\left[
\left\|
W \odot (\hat{w}_0 - w_0)
\right\|_1
\right],
\end{equation}
where $W$ applies higher weights near lesion boundaries and moderates diagonal HH amplification.

To stabilize brightness and prevent drift in low-prevalence regimes, we introduce low-frequency moment regularization:
\begin{equation}
\mathcal{L}_{LL}
=
\lambda_\mu \|\mu_{LL}^{pred}-\mu_{LL}^{tgt}\|_2^2
+
\lambda_\sigma \|\log \sigma_{LL}^{pred}-\log \sigma_{LL}^{tgt}\|_2^2.
\end{equation}

High-frequency variance control ensures texture fidelity without noise amplification:
\begin{equation}
\mathcal{L}_{HF}
=
\sum_{b \in \{LH,HL,HH\}}
\lambda_b
\|\log \sigma_b^{pred}-\log \sigma_b^{tgt}\|_2^2.
\end{equation}

Finally, mild pixel-space auxiliary constraints encourage edge alignment and prevent intensity saturation in lesion and peri-lesional regions.

The overall objective is:
\begin{equation}
\mathcal{L}_{total}
=
\mathcal{L}_{wavelet}
+
\mathcal{L}_{LL}
+
\mathcal{L}_{HF}
+
\mathcal{L}_{aux}.
\end{equation}

This formulation enables interpretable control over structure, detail, and brightness while maintaining computational efficiency.

\subsubsection{Structured Classifier-Free Guidance}

To disentangle lesion conditioning from anatomical context, we adopt structured classifier-free guidance. We compute three forward passes: (i) unconditional, (ii) mask-only, and (iii) mask+neighbor conditioned predictions. The final estimate is:

\begin{align}
\hat{w}_0^{SALIENT}
&=
f_\theta(x_t,t,\varnothing)
\\
&+
s_{mask}
\left[
f_\theta(x_t,t,c_{mask})
-
f_\theta(x_t,t,\varnothing)
\right]
\\
&+
s_{nei}(t)
\left[
f_\theta(x_t,t,c_{mask+nei})
-
f_\theta(x_t,t,c_{mask})
\right].
\end{align}

The neighbor guidance scale $s_{nei}(t)$ decays over time, encouraging global anatomical coherence early in diffusion and lesion-focused refinement in later steps.

This structured guidance allows SALIENT to sample morphologically diverse lesions while preserving anatomical plausibility.

\subsection{3D VAE for Volumetric Lesion Mask Generation}

To introduce mask-guided generative diversity beyond the limited observed positive set, we train a 3D variational autoencoder (MaskVAE3D) on volumetric lesion masks. Contiguous axial mask slices are stacked and resampled to a fixed size, forming spatially coherent binary volumes.

The encoder maps each volume to a low-dimensional latent representation, and the decoder reconstructs anatomically plausible lesion masks. Reconstruction combines Dice and boundary-weighted losses to preserve topology and surface fidelity. A KL regularization term with free-bits stabilization prevents latent collapse.

At inference, latent codes are sampled from a Gaussian prior and decoded into volumetric masks, which are sliced to provide diverse conditioning inputs for SALIENT.

\subsection{Semi-Supervised Segmentation for Paired Masks}

To obtain slice-aligned lesion masks for synthetic CT images, we apply a semi-supervised segmentation model based on Uncertainty-aware Cross-Model Training (UCMT)\cite{shen2023cotraining}. 

UCMT uses two student networks with an exponential moving average (EMA) teacher, combining supervised Dice-style loss on labeled real slices with cross-pseudo supervision and uncertainty-guided mixing on unlabeled data. 

After training, the frozen EMA teacher is applied to synthetic slices to produce binary lesion masks. These pseudo-labels provide geometrically consistent CT–mask pairs for downstream mask-guided classification.

\subsection{Mask-Guided Detection Evaluation}

We evaluate whether SALIENT-generated paired CT–mask samples improve downstream detection under severe class imbalance. We train a mask-guided slice-level classifier and aggregate slice evidence for subject-level prediction.

We adopt a ResNet-50 backbone augmented with two lightweight mask-guided attention (MGA) blocks inserted at intermediate and deep feature stages, following prior mask-guided attention paradigms \cite{li2018tell, yan2024enhanced}. Given an input slice (or 2.5D triplet), each MGA block produces a spatial attention map encouraged to align with the lesion mask during training.

Classification is optimized using focal loss to address class imbalance. In addition, we apply an attention-alignment loss (mean-squared error) between normalized attention maps and resized lesion masks to enforce spatial accountability. The overall objective combines focal and attention-alignment losses, weighted by a balancing coefficient. 

For patient-level prediction, we extract slice-level embeddings from the trained backbone and aggregate them using a lightweight Transformer encoder with positional embeddings. Following the Embedded Vision Transformer (EViT) \cite{islam2024seeking} paradigm, token aggregation emphasizes informative slices before producing a subject-level probability. The same aggregation scheme is used across all ablations to isolate the effect of synthetic paired augmentation.

\section{Experiments}

\subsection{Experimental Setting}

In this study, SALIENT was trained using AdamW with cosine learning-rate decay and exponential moving average (EMA) stabilization. Diffusion employed a cosine noise schedule and wavelet-domain band weighting to balance global structure and high-frequency detail. Low-frequency stabilization, high-frequency variance control, and mild pixel-space auxiliary constraints were applied as described in Sec.~3. Full hyperparameter configurations are provided in the Supplement. MaskVAE3D was trained with AdamW and EMA stabilization using boundary-aware reconstruction losses and KL regularization with free-bits stabilization. Training details and loss weights are provided in the Supplement. UCMT employed DeepLabv3+ backbones with supervised Dice-style loss and cross-pseudo supervision under uncertainty-aware mixing. Models were trained using AdamW with cosine decay. Full training parameters are included in the Supplement. The mask-guided ResNet50 backbone was trained with focal loss and attention-alignment supervision. To address imbalance, all positive slices were included per epoch with balanced negative sampling. Optimization used AdamW with cosine decay. Detailed settings are provided in the Supplement. For subject-level prediction, we extracted per-slice 2048-D features from the frozen slice-level backbone. Sequences were ordered by slice index and padded/truncated to length 500. A lightweight Transformer encoder (4 layers, 8 heads, embedding dimension 512) processed slice tokens with learned positional encodings. Training used AdamW (learning rate $10^{-4}$), cosine decay for 200 epochs, and balanced subject-level sampling.

\subsection{Generation Results}

We compare SALIENT against a pixel-space MedDDPM\cite{dorjsembe2024conditional} baseline trained with the same VAE mask generation and semi-supervised segmentation pipeline. The baseline operates directly in pixel space using 2D or 2.5D denoising, while SALIENT performs diffusion over discrete wavelet coefficients.

\subsubsection{Qualitative Comparison.}
\begin{figure*}[t]
    \centering
    \includegraphics[width=\textwidth]{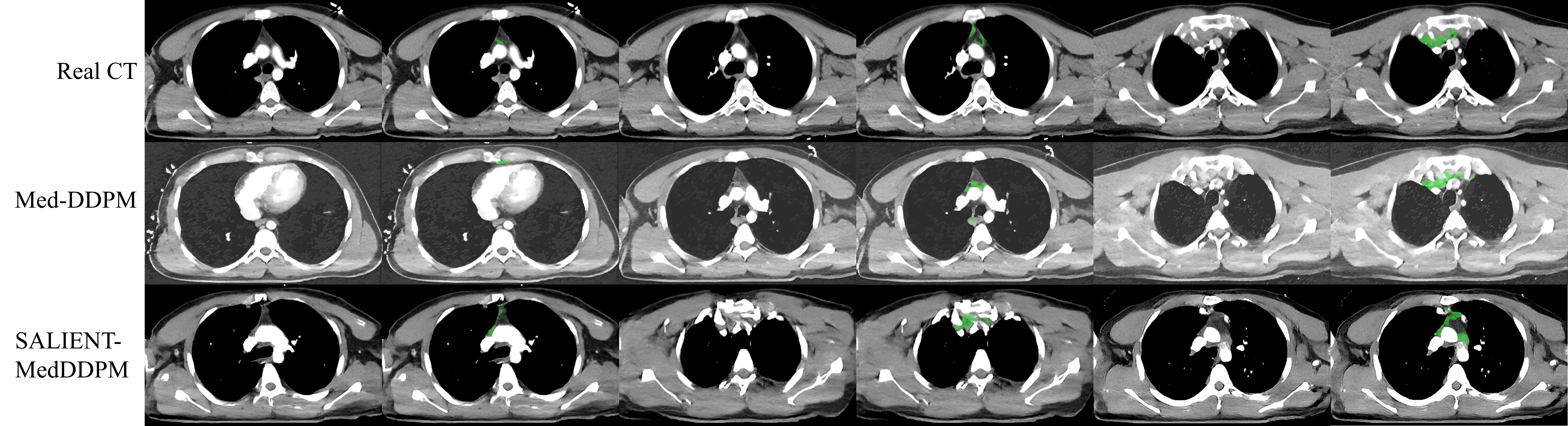}
    \caption{
    Comparison of synthetic CT slices generated by pixel-space MedDDPM and wavelet-domain SALIENT. 
    SALIENT produces sharper anatomical boundaries, reduced high-frequency noise, and improved contrast stability in mediastinal regions.
    }
    \label{fig:generation_result}
\end{figure*}

Figure~\ref{fig:generation_result} shows representative synthetic CT slices generated by MedDDPM and SALIENT alongside corresponding masks. MedDDPM outputs exhibit visible high-frequency noise, localized brightness amplification near the mediastinum, and occasional structural oversmoothing. In contrast, SALIENT produces sharper vascular boundaries, improved soft-tissue contrast, and more anatomically coherent lung–mediastinum interfaces. Lesion regions remain consistent with conditioning masks without introducing peripheral artifacts.

These qualitative differences are consistent across subjects and slice positions, particularly in challenging superior mediastinal regions where small hematomas occupy a limited fraction of the field of view. A board-certified radiologist performed blinded grading of synthetic CT slices containing mediastinal hematoma from 20 randomly selected cases using a 5-point Likert scale (1=poor, 5=near-indistinguishable from real). Graded criteria included global structural realism, lesion plausibility, lesion–background integration, high-frequency artifact presence (reverse scored), brightness and contrast realism, and mask fidelity (definitions in Appendix). SALIENT demonstrated higher brightness and contrast realism, improved lesion–background integration, fewer high-frequency artifacts, and markedly superior mask fidelity compared to pixel-space MedDDPM, with modest differences in global structural realism. These qualitative trends were consistent with quantitative segmentation fidelity (Dice: 0.72±0.24 vs 0.27±0.16) and aligned with the effects of SALIENT augmentation on precision in downstream detection dose-response experiments.

\subsubsection{Objective Realism Metrics.}
We quantify synthesis quality using multi-scale structural similarity (MS-SSIM) and Fréchet Inception Distance (FID). SALIENT improves MS-SSIM from 0.63 to 0.83 and reduces FID from 118.4 to 46.5:

\begin{center}
\begin{tabular}{lcc}
\hline
Method & MS-SSIM $\uparrow$ & FID $\downarrow$ \\
\hline
MedDDPM (pixel-space) & 0.63 & 118.4 \\
SALIENT (wavelet-domain) & 0.83 & 46.5 \\
\hline
\end{tabular}
\end{center}

The increase in MS-SSIM indicates improved structural fidelity across scales, while the substantial FID reduction reflects better alignment with real CT feature distributions.

\subsubsection{Wavelet-Band Analysis.}
\begin{figure*}[t]
    \centering
    \includegraphics[width=\textwidth]{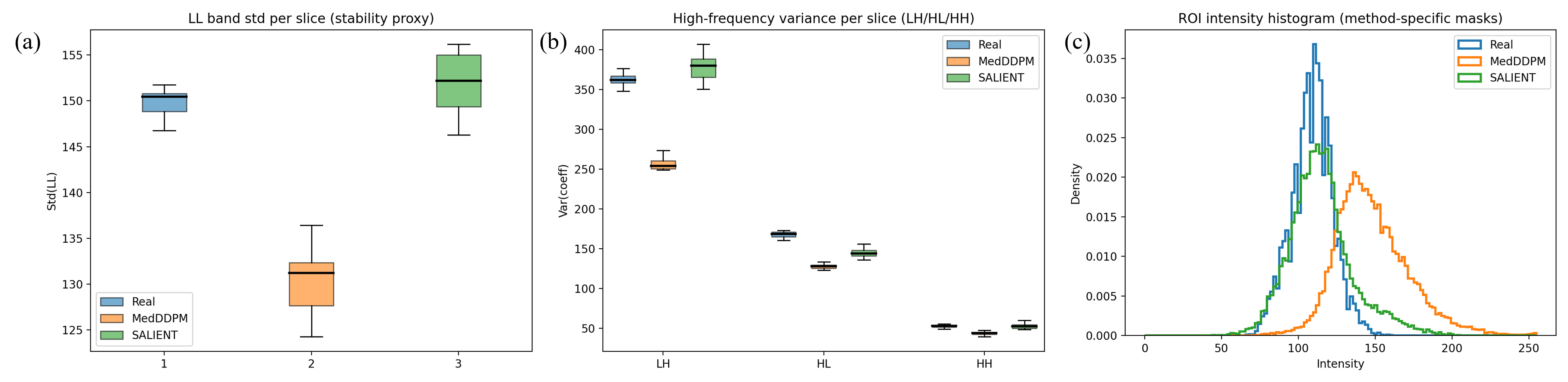}
    \caption{
    Quantitative frequency and intensity comparison between Real CT, pixel-space MedDDPM, and SALIENT. 
    (Left) Per-slice LL standard deviation. 
    (Middle) High-frequency variance per slice (LH/HL/HH). 
    (Right) ROI intensity histograms using method-specific masks.
    }
    \label{fig:per_band_stat}
\end{figure*}

We analyzed energy distributions across the four wavelet bands (LL, LH, HL, HH). Pixel-space MedDDPM exhibits unstable low-frequency brightness shifts and excessive high-frequency variance. In contrast, SALIENT’s band-wise diffusion with explicit LL/HF regularization produces energy profiles that more closely match real mediastinal hematoma data. As shown in Fig.~\ref{fig:per_band_stat}, the LL-band standard deviation indicates that MedDDPM underestimates global contrast and displays greater variability, consistent with brightness drift. SALIENT restores LL variability toward the real CT distribution, improving low-frequency stability. For high-frequency bands (LH, HL, HH), MedDDPM shows directional imbalance and variance distortion, indicating noisy or overly smoothed textures. SALIENT aligns more closely with real variance across detail bands, preserving anisotropic edge structure without amplifying HH artifacts. ROI intensity histograms further demonstrate that MedDDPM shifts lesion intensities toward higher values, with broader tails, whereas SALIENT maintains contrast distributions closer to those of real CT while avoiding saturation.

\subsubsection{Computational Efficiency.}
Wavelet-domain modeling also improves efficiency. SALIENT achieves approximately $4\times$ faster training relative to 2.5D MedDDPM, and $28\times$ speedup compared to full 3D MedDDPM (single NVIDIA H100), while preserving $512\times512$ in-plane resolution. This enables practical high-resolution synthesis without the computational burden of volumetric diffusion.

Overall, these results indicate that wavelet-domain diffusion with structured mask guidance yields sharper, more stable, and more computationally efficient synthesis than pixel-space MedDDPM.

\subsection{Detection Results}
\label{sec:cls_results}

We evaluate whether SALIENT-generated paired CT--mask samples provide \emph{functional} benefit in downstream detection, beyond perceptual realism. Using the mask-guided ResNet50 backbone, we report \textbf{subject-level} performance under severe class imbalance by controlling test prevalence (1--5\%) and varying the \emph{synthetic-to-real} augmentation ratio in training. Performance is summarized using AUPRC (primary under imbalance) and AUROC.

\subsubsection{Dose-response under synthetic augmentation.}
Table~\ref{tab:dose_response} reports the \emph{dose-response} of SALIENT augmentation (ratios $0\times$ to $10\times$). With a labeled seed of $n{=}50$ positive studies, SALIENT exhibits a consistent \textbf{therapeutic} regime at \textbf{$2\times$} synthetic augmentation across all prevalences, improving AUPRC by $\approx$0.05--0.06 absolute. When the labeled seed is reduced to $n{=}25$, the therapeutic dose shifts rightward to \textbf{$4\times$}, with substantially larger AUPRC gains (up to $\approx$0.12 at 1\% prevalence). Notably, AUROC remains high throughout, indicating that the primary effect of SALIENT augmentation is \emph{precision rescue} (AUPRC) rather than trivially inflating separability.

\begin{table}[t]
\centering
\small
\setlength{\tabcolsep}{4.2pt}
\begin{tabular}{c|c|c|c|c|c}
\hline
Seed & Prev. & AUPRC ($0\times$) & Best AUPRC & Opt. dose & $\Delta$AUPRC \\
\hline
\multirow{5}{*}{$n{=}50$}
& 1\% & 0.8809 & \textbf{0.9414} & $2\times$ & +0.0605 \\
& 2\% & 0.9062 & \textbf{0.9677} & $2\times$ & +0.0615 \\
& 3\% & 0.9110 & \textbf{0.9684} & $2\times$ & +0.0574 \\
& 4\% & 0.9215 & \textbf{0.9745} & $2\times$ & +0.0530 \\
& 5\% & 0.9223 & \textbf{0.9745} & $2\times$ & +0.0522 \\
\hline
\multirow{5}{*}{$n{=}25$}
& 1\% & 0.8169 & \textbf{0.9408} & $4\times$ & +0.1239 \\
& 2\% & 0.8776 & \textbf{0.9698} & $4\times$ & +0.0922 \\
& 3\% & 0.8803 & \textbf{0.9736} & $4\times$ & +0.0933 \\
& 4\% & 0.8831 & \textbf{0.9826} & $4\times$ & +0.0995 \\
& 5\% & 0.8843 & \textbf{0.9826} & $4\times$ & +0.0983 \\
\hline
\end{tabular}
\vspace{-0.25em}
\caption{\textbf{Subject-level dose-response} of SALIENT augmentation. ``Opt. dose'' denotes the synthetic-to-real ratio achieving the best AUPRC for each prevalence. SALIENT yields a stable therapeutic dose of $2\times$ at $n{=}50$ and a right-shift to $4\times$ at $n{=}25$, with larger gains in the low-label regime.}
\label{tab:dose_response}
\vspace{-0.5em}
\end{table}

To isolate the contribution of paired mask-conditioning, we compare against training without mask guidance (same real/synthetic sampling protocol). At 1\% prevalence ($n=50$), the baseline model without augmentation achieves an AUPRC of 0.8809. Synthetic augmentation without mask guidance fails to improve performance (best $\approx$0.8408 across ratios). In contrast, mask-guided SALIENT augmentation increases AUPRC to 0.9414 at $2\times$. These results suggest that performance improvements stem from paired image–mask supervision rather than synthetic image quantity alone.

We further stratified performance by target-to-volume ratio (TVR) at local (4\%) prevalence to assess robustness under within-patient signal dilution. SALIENT augmentation produced the largest AUPRC gains in the small-TVR regime (+0.1103), followed by large-TVR (+0.0832) and middle-TVR (+0.0764). In contrast, AUROC gains were modest (−0.0005, +0.0028, +0.0129 for small, middle, and large TVR, respectively), indicating that SALIENT primarily improves precision rather than ranking separability. These findings suggest that paired wavelet-domain augmentation is particularly effective when the lesion signal is diluted by low target-to-volume ratios.

\subsubsection{Qualitative evidence via saliency alignment.}
\begin{figure}[t]
    \centering
    \includegraphics[width=\linewidth]{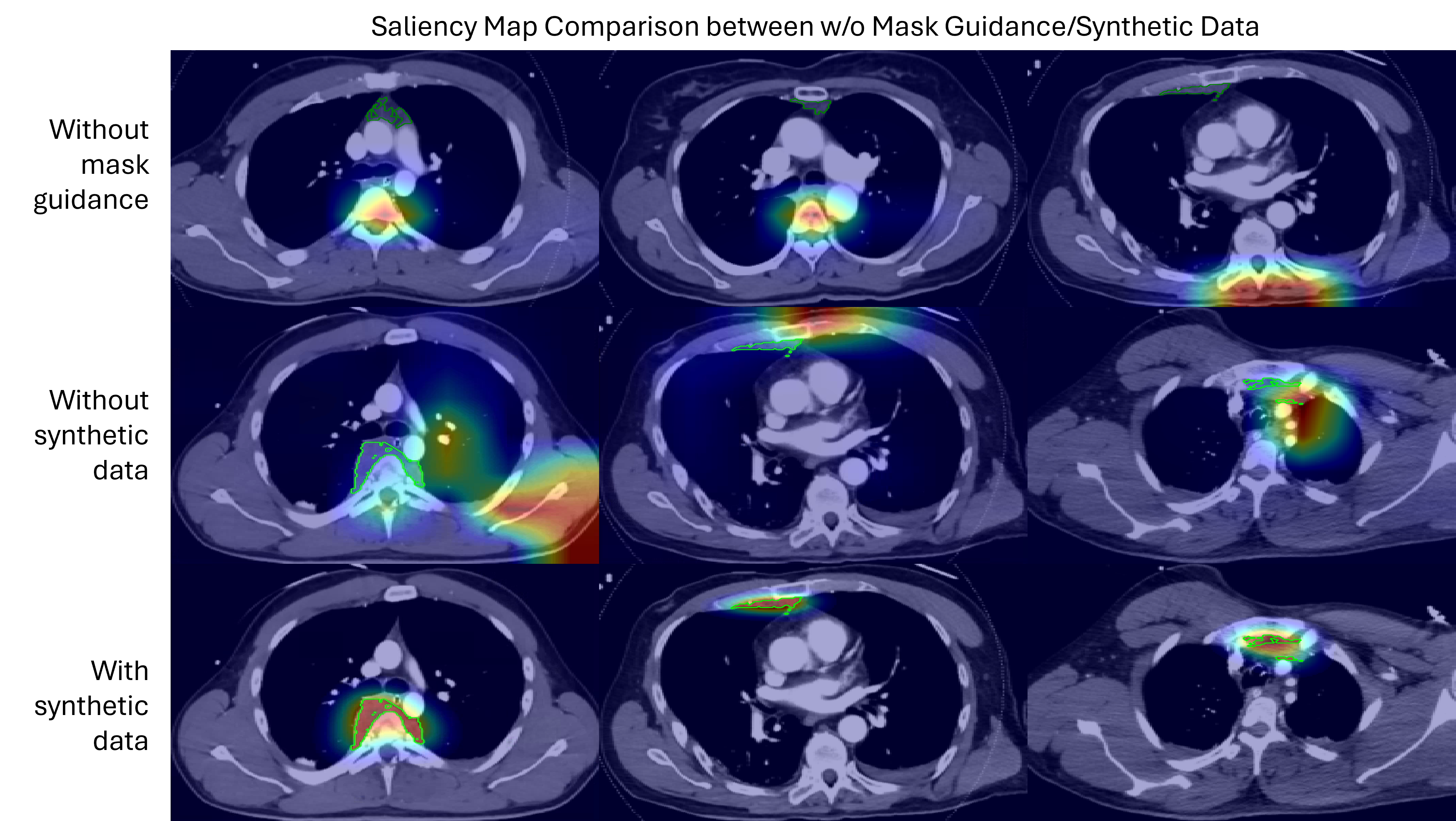}
    \vspace{-0.4em}
    \caption{\textbf{Saliency alignment improves with paired augmentation.}
    Top: training without mask guidance; middle: without synthetic data; bottom: with SALIENT paired CT--mask augmentation. SALIENT encourages lesion-focused evidence and reduces spurious activations on irrelevant anatomy.}
    \label{fig:saliency_map}
    \vspace{-0.6em}
\end{figure}

Figure~\ref{fig:saliency_map} visualizes saliency maps under three training conditions: (i) without mask guidance, (ii) without synthetic data, and (iii) with SALIENT paired augmentation. Without masks or synthetic pairing, saliency frequently concentrates on irrelevant structures (e.g., body wall), suggesting shortcut learning. In contrast, SALIENT paired augmentation increases overlap between saliency and the lesion region, supporting the claim that wavelet-domain diffusion enables clinically meaningful augmentation that improves where the detector attends, not only its aggregate metrics.

Across prevalences (1--5\%) and labeled seed sizes ($n{=}25,50$), SALIENT exhibits a reproducible augmentation dose-response with a clear therapeutic regime. The systematic right-shift in optimal augmentation under reduced label availability suggests that higher synthetic ratios may be beneficial in low-label regimes, provided mask guidance preserves spatial accountability.

\section{Conclusion}

We presented SALIENT, a mask-conditioned wavelet-domain diffusion framework for controllable CT augmentation under extreme class imbalance. By replacing pixel-space denoising with band-wise diffusion over discrete wavelet coefficients (LL/LH/HL/HH), SALIENT explicitly separates global brightness structure from high-frequency edge and texture components. This structured frequency decomposition enables stable optimization with target–background regularization, improved artifact control, and substantially reduced computational cost compared to volumetric diffusion.

Across mediastinal hematoma detection, SALIENT demonstrated consistent improvements in both perceptual realism and functional utility. Wavelet-domain modeling increased MS-SSIM and reduced FID while preserving high in-plane resolution with 4× faster training relative to 2.5D pixel-space diffusion. More importantly, paired CT–mask synthesis translated into reproducible downstream precision gains under severe class imbalance. We observed a stable augmentation dose-response: with sufficient labeled data ($n=50$), the therapeutic regime occurs at $2\times$ synthetic augmentation, while in low-label settings ($n=25$), the optimal dose shifts to $4\times$. Mask guidance proved essential for accountable gains, improving both saliency alignment and AUPRC relative to unpaired synthetic augmentation. Notably, gains were strongest in low prevalence and target-to-volume ratio regimes, highlighting SALIENT’s ability to counteract prevalence- and within-patient signal dilution in precision-sensitive settings through combined lesion-mask augmentation.

These findings suggest that frequency-aware diffusion provides a practical mechanism for controllable precision rescue in long-tail detection problems. By enabling explicit regulation of brightness and high-frequency structure, SALIENT transforms synthetic data from a heuristic augmentation strategy into a tunable component of the training pipeline. Future work will explore extensions to additional long-tail CT detection tasks involving irregular, spatially heterogeneous, and multiscale lesion morphology.

%
%
\bibliographystyle{splncs04}
\bibliography{main}
\end{document}